\documentclass{JHEP3}
\usepackage{amssymb}
\usepackage{amsfonts}
\usepackage{amsbsy}
\def\be{\begin{equation}}
\def\ee{\end{equation}}
\def\ba{\begin{array}}
\def\ea{\end{array}}
\newcommand{\bea}{\begin{eqnarray}}
\newcommand{\eea}{\end{eqnarray}}

\def\Vol{\mbox{Vol }}

\def\IZ{\mathbb{Z}}

\def\CR {{\cal R}}

\def\CL {{\cal L}}

\def\CO {{\cal O}}

\def\CA{{\cal A}}
\def\CK{{\cal K}}

\def\vev#1{\langle{#1}\rangle}

\newcommand{\eq}[1]{Eq.~(\ref{eq:#1})}

\renewcommand{\Im}{{\rm Im }}

\def\one{{\hbox{ 1\kern-.8mm l}}}

\def\tr{{\rm tr\,}}

\def\ba{\bar{a}}

\title{On $D=5$ super Yang-Mills theory and
$(2,0)$ theory}

\author{Michael R. Douglas$^{1,\&}$\\
$^1$Simons Center for Geometry and Physics\\
Stony Brook University\\
Stony Brook, NY 11794 USA\\
\\
$^\&$I.H.E.S., Le Bois-Marie, Bures-sur-Yvette, 91440 France\\
{\tt douglas@max2.physics.sunysb.edu}}

\abstract{
We discuss how $D=5$ maximally supersymmetric Yang-Mills theory (MSYM) might be
used to study or even to define the $(2,0)$ theory in six dimensions.  It is known that the
compactification of $(2,0)$ theory on a circle leads to $D=5$ MSYM.
A variety of arguments suggest that the relation can be reversed, and that all of the degrees
of freedom of $(2,0)$ theory are already present in $D=5$ MSYM.  
If so, this relation should have consequences for 
$D=5$ SYM perturbation theory.  We explore whether it might imply
 all orders finiteness, or else an unusual relation between the cutoff and the gauge coupling.
 S-duality of the reduction to $D=4$ may provide nonperturbative constraints or tests of 
 these options.
}

\begin{document}

\section{Introduction}

Among the many discoveries of the second superstring revolution which we have not
yet fully understood, is the existence of interacting local quantum field theories
in spacetime dimensions $4<D\le 6$ \cite{Witten:1995zh}.  
So far as we know, such theories must be supersymmetric,
and the most intriguing examples are the so-called $(2,0)$ theories,
with $2$ chiral supersymmetries ($16$ supercharges) and superconformal invariance
in $D=6$.  These theories 
are classified by a choice of simply laced Dynkin diagram (so, $A_n$, $D_n$,
$E_6$, $E_7$, $E_8$ and their direct sums) and can be obtained
as low energy decoupling limits of IIB strings compactified on $K3$.  The $A_n$ series
can also be obtained as the low energy limit of the world-sheet theory of $n+1$ coincident
M5 branes.  

No action is known for these theories, and according to the string/M theory definition
they have no dimensionless parameters in which to expand, suggesting that 
perturbative computations would not be possible even given an action.  Thus, not much
is known about these theories beyond the predictions of supersymmetry and duality arguments.

In this note, we discuss the possibility of using the compactification of $(2,0)$ theory on a circle (or $S^1$),
which leads to $D=5$ maximally supersymmetric Yang-Mills theory (MSYM), to learn more about
both theories.  Following \cite{Seiberg:1996bd,Seiberg:1997ax},
we review the basics of this relation in section 2.
The main point which we need for our introductory discussion is that the (squared) gauge coupling,
which in $D=5$ has dimensions of length, is equal to the radius of the $S^1$ in
compactified $(2,0)$ theory,
\be
g_5^2 = R_6 .
\ee
Thus, the equivalence to $D=5$ SYM should be valid in the low energy limit $E<<1/g_5^2$,
with corrections controlled by the dimensionless parameter $g_5^2 E$.

Of course, the first problem in making sense of this relation is that 
by the usual power counting arguments, $D=5$ super Yang-Mills
is perturbatively nonrenormalizable.  Thus one
expects an infinite series of divergences requiring an infinite series of counterterms, and an
expansion with
little or no predictive power.  However, the situation is better in supersymmetric theories.
Superspace arguments preclude divergences at low orders, and in $D$-dimensional MSYM
the first divergences appear at $\ell$ loops with $D=4+6/\ell$ \cite{Bern:1998ug}.  And, 
as the technology for higher loop perturbation theory
has advanced, remarkable cancellations have been discovered 
which go beyond the predictions
of the superspace arguments.  Until a computation which could have diverged
consistently with these arguments is done, for example at $\ell=6$, it is hard to
evaluate the situation here.  But, at present it is not ruled out that $D=5$ MSYM is
perturbatively finite to all orders, just as is true for $D=4$.

There is a close analogy to the old question of to whether $N=8$ supergravity 
in space-time dimension $D=4$ is a perturbatively finite quantum field theory, or not.
Superspace arguments predict the first divergences at seven loops, which is still out
of reach of explicit verification.  However, the same arguments predict other
divergences in higher dimensions, which recently have been shown to cancel 
at four loops \cite{Bern:2009kd}.
At present the systematics of these divergences is unclear; see
\cite{Elvang:2010jv,Bjornsson:2010wm,Bossard:2010bd} for recent discussions.

Suppose that $D=4$ supergravity were finite to all orders.
Then the mystery would be far deeper, as there 
is no known candidate for a fully consistent $D=4$ quantum field theory ({\it i.e.}, defined
nonperturbatively and at all energy scales) for which the scattering amplitudes 
would have $N=8$ supergravity as an asymptotic expansion.  

Let us briefly recall why this is.
As in almost all interacting quantum field theories
with a finite number of fields, the perturbative expansion is expected to have zero
radius of convergence (the $\ell$ loop term behaves as $\ell !$) and by itself does not define
a theory.  One must propose either a resummation prescription, or (better) a physical picture
for how the exact amplitudes behave in the UV, or (ideally)
a way to obtain the theory as the low energy limit of some quantum theory which is known to exist --
in other words a UV completion -- to claim that a fully consistent QFT exists.  

This is especially important and nontrivial for a perturbatively nonrenormalizable theory, 
such as quantum gravity in $D>2$ and quantum Yang-Mills in $D>4$.  While
the formal loop expansion parameter $g^2$ has dimensions of
length to a positive power, the actual behavior of perturbation theory is controlled by a
dimensionless effective coupling.  Consider a scattering amplitude at energy $E$; this will be
$g^2 E^\alpha$ for some $\alpha>0$, so at high energies perturbation theory must
break down.  

At present, the only candidate for a
UV completion of quantum gravity is compactified superstring theory, which is perturbatively 
finite for reasons having no relation to field theoretic
perturbation theory.  It is not clear what significance perturbative
finiteness of supergravity
would have in this context; superstring theory would be finite with or without it.  See \cite{Green:2007zzb}
for a recent discussion.

By contrast, we know a UV completion of
$D=5$ maximally supersymmetric Yang-Mills theory, namely compactified $(2,0)$ theory, so
the question of whether it is perturbatively finite or not should
have direct consequences for this theory.  In this note we begin to explore this idea.

Let us begin with a temptingly simple argument as to why $D=5$ MSYM
must be perturbatively finite.  Suppose it were not.  In this case, we would 
need to cut off the theory at a scale $\Lambda$.  This would introduce a 
new dimensionless parameter $g_5^2 \Lambda$.  On the other hand, the $(2,0)$ theory does not
have dimensionless parameters, leading to a contradiction.  

To make this a bit more precise, the assumption is that the low energy behavior of compactified
$(2,0)$ amplitudes looks like $D=5$ MSYM below some adjustable scale $\Lambda$, and
is different above this scale, perhaps because of additional particles, perhaps because the
interaction has a form factor, or for some other reason.  It does not really matter  
what the nature of the cutoff is for this argument, only that the scale $\Lambda$ can be extracted from the 
compactified $(2,0)$ amplitudes, and is adjustable independently from $g_5^2$.

While this argument has loopholes which we will explain shortly, it should be taken
seriously, as this is how many
analogous examples work.  For example, while perturbative string theory cuts off the divergences
of quantum gravity, the cutoff is not at the Planck scale
but rather at the string scale; the ratio of these parameters indeed defines a dimensionless
parameter, the string coupling.  
As an even more basic example, the UV divergences of the Fermi theory of weak
interactions are cutoff at the W and Z boson masses; comparing these to $G_{Fermi}$ defines 
new dimensionless parameters, the Standard Model gauge couplings.

Not all nonrenormalizable theories work this way.
For example, eleven dimensional M theory has a single
dimensionful scale $M_{p11}$, which determines both the gravitational coupling and
the scale at which perturbation theory breaks down.  In this example, it seems
natural that the two scales should be related.  The same might be true for $D=5$ MSYM,
evading the contradiction.

\subsection{Cutoff and extra states}

An important difference between M theory and $(2,0)$ theory is that in the latter case,
the underlying UV theory is scale invariant.  Thus, the cutoff $\Lambda$ only appears upon
compactification.  On $S^1$, this defines a single new scale $R_6=g_5^2$, so the only option is
that 
\be \label{eq:cutoff}
\Lambda=\frac{c}{g_5^2},
\ee
where $c$ is an order one constant, whose
precise definition would depend on the renormalization scheme in $D=5$ SYM.
Conceivably, $c$ might not be constant; it might turn out to depend on loop order
or other details of the diagrams being cutoff.  Our arguments will not depend on its
value, and often we will set $c=1$ below.

We should think of \eq{cutoff} as modeling the effect of some precise cutoff
provided by the compactified $(2,0)$ theory.  Since this theory 
has KK modes at the scale $1/R_6=1/g_5^2$, the 
simplest hypothesis is that adding these modes cuts off the UV divergences.

In fact, these KK modes can already be seen in $D=5$ MSYM.  As we review in section 2,
they are the self-dual solutions which would have been instantons in $D=4$.  
In fact, the
other BPS states of $(2,0)$ theory can also be seen in $D=5$ -- for example, the solutions
which would have been monopoles in $D=4$ become strings, which are precisely the
strings which are supposed to arise from M2-branes stretched between M5-branes in this
construction of $(2,0)$ theory.  

This is all consistent with the hypothesis that {the new degrees
of freedom which are required to define the UV completion, are {already present} in the nonperturbative
physics of the $D=5$ theory}.\footnote{In Lambert {\it et al} \cite{Lambert}, which appeared after this work was
substantially completed, this idea is suggested independently, and new evidence is given.
They also believe this suggests that $D=5$ MSYM could be perturbatively finite.}
If so, a nonperturbative formulation of $D=5$ MSYM would {\it ipso facto} be a formulation of
the $(2,0)$ theory, by taking the infinite coupling limit.

Although we do not have a nonperturbative formulation of $D=5$ MSYM, one can nevertheless
get nonperturbative information about a quantum theory by close examination of the perturbative
expansion.  For example, the $(2g)!$ asymptotics of closed string perturbation theory 
suggested the importance of objects with mass $1/g_s$, before the realization that these were D-branes \cite{Shenker:1995xq}.
The general principle \cite{Lipatov:1976ny,ZJ} is that the large order behavior of perturbation theory
is controlled by configurations in which the fields are `large,' and more specifically by certain
classical solutions.   For example, in bosonic $D=4$ gauge theory, the asymptotic behavior has been
argued to be controlled by a certain complex instanton \cite{Lipatov:1978en}.  An interesting application of these
ideas to multiparticle production in $D=4$ MSYM appears in \cite{Kharzeev:2009pa}.

Conversely, if one tries to define a field theory nonperturbatively by resumming the perturbative expansion,
one often finds that the resummation is ambiguous in a way that is resolved by adding in the effects of the
classical solution.  Mathematically, this is the case in which the Borel transform has singularities on the positive
real axis.  While in practice it is very difficult to make such resummation procedures precise, such an argument
can give important clues about nonperturbative physics.

As we explain in section 2, the large order behavior of $D=5$ scattering amplitudes, at least if they behave
as in other field theories, suggests that it is somehow controlled by the self-dual particle solution.
This suggests that a close analysis of the physics of this solution might be key to understanding $(2,0)$ theory,
as has been hinted in previous works \cite{Aharony:1997an,Collie:2009iz}.

To conclude this subsection, we should point out that one could make different hypotheses about the relation
between $D=5$ SYM and $(2,0)$ theory, in which the $(2,0)$ theory has additional or different microscopic 
degrees of freedom.
A relation like \eq{cutoff} could arise naturally in such an effective low energy Lagrangian, for
example the chiral Lagrangian which describes pions works this way \cite{Weinberg:1978kz}.  Thus it is important
to look for evidence distinguishing the scenario described here, from other possibiities.

\subsection{Consequences of divergences for reduction to $D=4$}

How can we test the ideas we just described, and more specifically \eq{cutoff} ?  Since we cannot
do the necessary nonperturbative computations directly in $D=5$ and $D=6$, let us instead consider
its consequences in $D=4$.  

Thus,
let us consider the compactification $D=5$ to $D=4$ on a circle,
or equivalently compactify the $(2,0)$ theory on $T^2$ with a flat metric.  This leads to
$N=4$ SYM, as we will review in section 3, but let us again state the basics for purposes of
our introduction.  Consider
the special case of a rectangular torus, then the radii of the torus are related to the $D=4$ gauge 
coupling as $R_5=L/g_4$ and $R_6=g_4 L$, where $L$ is a parameter with dimensions of length.
Taking the limit $L\rightarrow 0$ at fixed $g_4$, we recover the finite $D=4$ theory, while
taking small but finite $L$ produces corrections to the $D=4$ amplitudes.  These will be described
by irrelevant operators such as $L^4 \tr F^4$, $L^6 \tr D^2F^4$ and so on,
whose coefficients are each a function of $g_4$.

Now, if there are divergences in $D=5$, we argued earlier that
there must be a nonperturbative consistency condition relating $\Lambda$
and $g_5^2$.
A natural source of such a nonperturbative consistency condition is S-duality of $D=4$ MSYM.
Specifically, the scattering amplitudes of states which map into themselves under S-duality (the $U(1)^r$ gauge multiplets)
must be invariant under S-duality, including any corrections controlled by $L$. 
For example, the scattering of four photons gets a one loop contribution from charged $D=5$ KK gauge bosons.
Extending this to a sum including 6d KK modes will suggest an S-duality invariant extension \eq{S-dual-amp}
of this amplitude.  The resulting $\tr F^4$ term only receives contributions at one loop 
and nonperturbatively, and is finite.
But it seems reasonable to believe that, if we could compute a higher
correction of this type from $D=5$, and if the computation required cutting off UV divergences,
it would be invariant under S-duality only for a very specific cutoff prescription, and for a unique value of $\Lambda$.

To understand the $\Lambda$ dependence of these corrections, as we will discuss in section 3,
one must address the following issue.
An implicit assumption made in all work on these theories, and supported
by the results, is that in a situation like this, the $L\rightarrow 0$ limit is regular.  Suppose we have
a higher dimensional field theory, compactify it, and take a limit which sends the KK and other higher
dimensional states off to infinite mass; then the corrections to the lower dimensional field theory go
smoothly to zero.  Here, as we take $L\rightarrow 0$, the corrections must vanish as positive powers
of $L$, for any fixed $g_4$.

While this is evident at the classical level, and even with finite quantum corrections, 
it can be spoiled by UV divergences in the higher dimensional theory.  Let us consider the case at hand, 
thought of as a compactification of $D=5$ SYM to $D=4$, and suppose that there were 
a $D=5$ perturbative UV divergence.  Since $D=4$ MSYM is finite, 
the compactified $D=5$ counterterms would come entirely from diagrams involving
$D=5$ KK states, which have masses between $1/R_5$ and the cutoff $\Lambda$.  Although as $L\rightarrow 0$
both of these scales become large, if $\Lambda >> 1/R_5$ (as it is here), these quantum effects will survive the limit.
Since in a nonrenormalizable theory one expects corrections which come as
positive powers of $\Lambda$, this could easily violate the regularity assumption.

In compactification of string theory, or of a (presumably) finite field theory such as $(2,0)$ theory, this
potential problem should be solved by whatever physics makes the theory finite, as this will
suppress the contributions of KK states with masses above the cutoff.  
In $D=5$ terms, this potential problem must
be solved by our assumption \eq{cutoff}.

Let us consider a $D=5$ UV divergent counterterm.  As we discuss in section 3, these can be estimated
by expanding the $D=5$ Feynman diagrams in sums over the 5d KK momentum.  They take the general
form $(\Lambda/M_5)^n$ or perhaps $\log \Lambda/M_5$, where $M_5=1/R_5$ is the KK scale.
We then take the counterterm and, granting that $\Lambda=1/g_5^2=1/R_6$, express it in $D=4$ terms.

Since $\Lambda/M_5=1/g_4^2$, this procedure introduces negative powers of $g_4$.  A UV correction with
an overall dependence $1/g_4^{2n}$ or $\log g_4$
would be very problematic, as it would become arbitrarily large at small $g_4$, and
at least naively would contradict the regularity assumption.
While one could escape the contradiction by postulating that higher order corrections dominate
this one, then the perturbative expansion would completely break down.  

Carrying this our more carefully in section 3,  we find that the UV corrections
all come with positive powers of $g_4$ and $L$, and in this sense are consistent 
with our interpretation
of the $D=5$ divergences as cut off in compactified $(2,0)$ theory.  Thus there seems to be no {\it a priori}
argument, at least from these considerations, that $D=5$ MSYM must be finite.  However, log UV divergences
leave an unusual signature in $D=4$.

This is as far as we have gotten with general arguments, and clearly the question of divergences 
in $D=5$ MSYM at a given (low)
loop order would be better settled by direct computation.  Whatever comes out, we believe we have made the point 
that the answer is quite relevant to the structure of $(2,0)$ theory.

\section{The $(2,0)$ theory and $D=5$ MSYM}

Let us review the basic properties of the $(2,0)$ theories \cite{Seiberg:1997ax}.
Their original definition was in terms of superstring theory.  One can
compactify the IIb string on a K3 surface, leading to a theory whose low energy limit
is a six-dimensional supergravity.  The K3 surface has metric moduli (volumes of two-cycles)
which can be taken to singular limits, classified by the simply laced (ADE) Dynkin diagrams.
In these limits, a new energy scale $\vev{\phi}$ appears, parametrically
lower than the string and Planck scales.  One can argue that the dynamics at this scale
decouples from supergravity, leaving a nontrivial interacting six-dimensional field theory.
Taking the limit $\vev{\phi}\rightarrow 0$, one obtains a scale invariant theory.

In six dimensions, $(2,0)$ supersymmetry is maximal (for a non-gravitational theory), and
there is a unique matter multiplet with this supersymmetry, containing a self-dual two-form
potential $B$, five scalars, and $4$ chiral fermions.  
The theory with a single matter multiplet can be realized as the low energy limit of the
world-volume theory of a five-brane in M theory.  Since the only parameter of the
five-brane theory is the eleven-dimensional Planck scale, in this limit there are no
adjustable parameters, dimensionful or dimensionless.   
The presence of a self-dual field makes
a covariant Lagrangian description tricky (see \cite{Pasti:1997gx}), but a noncovariant Lagrangian can
be written \cite{Aganagic:1997zq}; it is unique with no free parameters.  

One can also define the A series theories by taking the decoupling limit
of the world-volume theory of $n$ coincident five-branes in M theory.  By separating the
branes (going out of the Coulomb branch), and taking a further low energy limit, one obtains
a sum of the single multiplet theories we just described.  This supports the idea that the interacting
theories have no free parameters; and that the interaction is (in some sense) order one.
This fits with the fact that objects charged under self-dual gauge fields, satisfying
$$
H \equiv dB = *dB ,
$$
must satisfy the Dirac quantization condition simultaneously as electric and as magnetic
objects, forcing their charge to be order one.

\subsection{Reduction of the $(2,0)$ theory to $D=5$}

We now compactify $x^6$ on a circle of circumference $2\pi R_6$.
For reasons we explain shortly, we give this parameter several names:
\begin{equation} \label{eq:mdef}
\frac{g_{5}^2}{8\pi} \equiv \frac{1}{M_6} \equiv R_6 .
\end{equation}

While we have no action for the $(2,0)$ theory, as we discussed above, 
by turning on appropriate expectation
values $\Delta_{ij}\phi\equiv\phi_i- \phi_j \ne 0\forall i,j$ for the scalars
and taking a low energy limit, it reduces to a
sum of $D=6$ self-dual tensor theories. 
Applying standard dimensional reduction to these theories,
we find a sum of $D=5$ super-Maxwell actions, where the 5d
gauge field strength is $F_{\mu\nu}=H_{6\mu\nu}$.
The other components of $H$ are determined by the self-duality condition.
Then, given that the compactification  to $D=4$ is MSYM, we conclude that
the limit $\phi_i\rightarrow 0$ in $D=5$ should also be MSYM.

We next discuss some standard properties of the perturbative expansion
of MSYM in $D=5$.  Let us just for simplicity start from the
assumption that it is UV convergent, 
then by dimensional analysis and the standard asymptotics
of perturbation theory, the $\ell$ loop contribution to
a fixed angle scattering amplitude at energy $E$ is expected to grow as 
$$
{\cal A}_\ell \sim \ell ! \left(\frac{E}{M_6}\right)^\ell .
$$
Of course, there could be UV divergent terms.
Such a term will come with an additional factor of
$(\Lambda/E)^k$ (or perhaps logarithms).  To make the finiteness argument we stated
in the introduction, we need to show that this expansion can be written in terms
of $\Lambda/M_6$, and argue that it is plausible that terms in the expansion can be
reconstructed from the exact amplitudes.\footnote{
Given our ignorance of $(2,0)$ theory, one can even question whether an S-matrix can be
defined at all.  One way to argue that it can is to turn on linearly varying scalar vevs (in the
brane language, put the branes at small angles), to turn off the interactions in the asymptotic
region.  We thank Greg Moore for a discussion on this point.}
Now, the maximal possible UV divergence in a $D=5$ bosonic gauge
theory at $\ell$ loops\footnote{
This leaves out the vacuum energy, but this is zero for MSYM.}
appears in front of the $\tr F^2$ term and is $\Lambda^{\ell}$, so  already
in the bosonic case one can reformulate the series expansion as
$$
{\cal A}_\ell \sim \ell ! 
\sum_{k=0}^\ell \left(\frac{E}{M_6}\right)^{\ell-k} \left(\frac{\Lambda}{M_6}\right)^k.
$$
with all powers of $M_6$ in the denominator.
Of course, the UV divergences in the supersymmetric case are far milder, so the same is true.

This corresponds to an expansion
in positive powers of $g_5^2$, and thus one can isolate functions of the dimensionless parameter
$\Lambda/M_6$ at each order in such an expansion.
This is consistent with the idea that one can 
determine $\Lambda/M_6$ from an asymptotic expansion around $g_5=0$
of the (unknown) exact $D=5$ amplitudes, and that the terms in this expansion are the
terms computed in $D=5$ perturbation theory.  

Let us very briefly review what this means,
to make the point that, given certain hypotheses about the exact amplitudes, we could
determine the perturbative expansion and thus $\Lambda/M_6$.  Very generally in 
quantum theories, an amplitude $\CA(g)$ will be analytic in some wedge around the
origin $g=0$ containing the positive real axis.  Within this wedge, it has an
asymptotic expansion in $g$ such that
\begin{equation}
{\cal A} = \sum_{k=0}^{r-1} a_k g^k + {\cal R}_k(g)
\end{equation}
with a remainder term $\CR$ satisfying a bound
\begin{equation}  \label{eq:R-bound}
|{\cal R}_k(g)| \le C \sigma^r r! |g|^r
\end{equation}
uniformly in $r$ and $g$.  Because of this, by taking derivatives at suitably small
$g$, the terms $a_k$ in the expansion can be estimated to arbitrary accuracy.

While we do not know how to argue that exact compactified
$(2,0)$ theory amplitudes have the required analyticity,
if negative powers of $g_5$ had appeared, the claim would be evident nonsense, so we can at least
observe that a nontrivial consistency check has been passed, and continue on this assumption.
 
Granting it, if $D=5$ MSYM has perturbative UV divergences, these plausibly signal the
presence of a dimensionless parameter $\Lambda/M_6$ in the exact amplitudes.
Since the compactified $(2,0)$ theory has no such parameter, we have a reason to think
that $D=5$ MSYM is UV finite.  If not, there must be some
consistency condition which prevents us from varying the $(\Lambda/M_6)^k$ terms.
We now model this by setting all such terms to $1$.

Next, we can make a plausible guess for the behavior of the exact amplitude by imagining
that the expansion can be Borel resummed \cite{ZJ}, leading to
$$
{\cal A} \sim \exp -\frac{M_6}{E} .
$$
Now, such resummation arguments are not easy to make precise in practice,
as the physical effects of solitons and instantons tend to be associated with singularities
of the Borel transform on the positive real coupling axis, and the resummation
procedure becomes ambiguous beyond the one-instanton or one-soliton level  \cite{ZJ}.  Still,
many examples are known of theories in which the asymptotic behavior of 
perturbation theory is controlled by simple nonperturbative solutions,
so we can take this as suggesting an important nonperturbative role for a soliton 
associated with the mass $M_6$. 

\subsection{Instantonic particles}

An important clue to the UV completion of the 5d theory  \cite{Seiberg:1996bd} is the fact that the familiar
self-dual solution of Yang-Mills theory, satisfying $F=*F$, describes a particle
in $D=5$, which classically has mass
$M_6=8\pi/g_{5}^2$, as in Eq. (\ref{eq:mdef}).

These particles carry a conserved charge, with current
$$
J_6^\lambda = \epsilon^{\mu\nu\rho\sigma\lambda} \tr F_{\mu\nu} F_{\rho\sigma} 
$$
related to the instanton number in $D=4$.  It enters into a central charge of the
supersymmetry algebra and thus by BPS arguments, the particle mass
cannot receive quantum corrections.

From the compactified 6d point of view, this conserved charge is
the momentum in the sixth dimension.  This can be seen by reducing
the 6d expression for the appropriate component of the stress tensor,
$$
T_{06} = H_{0}^{ij} H_{6ij} = \epsilon^{0ijkl} F_{kl} F_{ij} .
$$

To summarize, $D=5$ super Yang-Mills indeed contains a particle at the mass scale $M_6$
at which perturbation theory breaks down, which can be identified with a Kaluza-Klein
mode of the $(2,0)$ theory.  It is an attractive hypothesis that adding
this particle to the $D=5$ theory, a soliton made up from the original fields of the theory,
gives us the full nonperturbative particle spectrum, and determines the nonperturbative
completion.  Of course, it is not immediately
obvious how to do this.  

In particular, compared to better understood solitons,
these solutions have rather peculiar properties.  For example, one of their bosonic
zero modes is a scale size $\rho$; classical solutions exist with arbitrary size.
Supersymmetry arguments imply that this mode will not be lifted in perturbation
theory, so apparently the energy-momentum of the solution can be arbitrarily widely
dispersed.  One might well question whether `particle'  is an appropriate name for
such a thing.  Possibly, as
suggested in \cite{Collie:2009iz}, this is a signal that these solutions
are bound states of several particles, the so-called `fractional instantons' or `partons'.

If we turn on scalar vevs, this scale invariance is broken, and (by analogy with
discussions of constrained instantons in $D=4$) the scale mode $\rho$ is confined to
small values.  One could then quantize it to get a more conventional looking particle.\footnote{
In \cite{Lambert}, this is demonstrated for the closely related dyonic solution.} 
In addition, a second soliton in the theory appears, the analog of the 't Hooft-Polyakov
monopole solution, which in $D=5$ is a string with tension $T=\vev{\Delta\phi} M_6$.
This string has  chiral
zero modes, with a very interesting anomaly cancellation mechanism discussed in
\cite{Boyarsky:2002ck}.  See \cite{Henningson:2004ix} for further analysis of the interpretation
of these objects in $(2,0)$ theory.

It would be interesting to further explore
the Coulomb phase $0\ne\Delta\phi<<M_6$, which exhibits the same UV puzzles, but
avoids many of the other puzzles of the nonabelian theory.

\subsection{Non simply laced gauge groups}

There are good string theory arguments that only the ADE theories exist in $D=6$.
Thus, to get the BCFG gauge groups in $D=5$, we must modify the dimensional reduction.
This can be done by positing twisted boundary conditions on the $S^1$.  Although we
do not have a concrete $D=6$ definition, presumably automorphisms of the Dynkin
diagram correspond to symmetries of the $(2,0)$ theory.  Thus one twists by such
an automorphism.  After reduction, this corresponds to 
twisting the gauge fields by an outer automorphism of the gauge group.  
It is known that the fixed points of such symmetries generate all of the non simply laced groups.
One might be able to test this idea by reducing to $D=4$ and making contact with 
the discussion of S-duality for BCFG groups in \cite{Kapustin:2006pk}.

\section{Reduction of the $(2,0)$ theory and $D=5$ MSYM to $D=4$}

Let us turn to $T^2$ compactification.  A flat metric on $T^2$ has
three real parameters: two radii $R_5$ and $R_6$, and a relative angle between the
A and B cycles.  We will group these into a dimensionless complex structure modulus $\tau$
which becomes the complexified $D=4$
gauge coupling $\tau=\theta/2\pi + 8\pi i/g_4^2$, and the volume 
$L^2=\Vol(T^2)$, a parameter with dimensions of length squared.

Our basic strategy will be to consider this theory as the compactification of $D=5$
SYM on a circle of radius $R_5$.  Thus, we will restrict attention to a rectangular $T^2$
with $\theta=0$.  Before doing so, let us briefly discuss how $\theta_{QCD}$ arises in
$D=4$, and its $D=5$ analog.  In the self-dual tensor theory, one has a term in the
the $D=6$ action $g^{56} H_{5\mu\nu} H_{6\mu\nu} $, which using self-duality
reduces to $\theta F\wedge F$.
In $D=5$, the metric components $g_{\mu 6}$
become a vector coupling to $J^\mu_6$.  Thus the $\theta$ angle shifts the
quantization of the momenta of the instantonic particles.  The $D=4$ instantons arise
as Euclidean winding configurations of these particles.

We now take $\theta=0$; then there is
an evident (geometric) symmetry under the interchange
$$
R_5 \leftrightarrow R_6 .
$$
Taking the low energy limit, this is the origin of S-duality in $D=4$.
Evidently it extends to an exact symmetry of the UV completion.

In $D=5$, this interchange corresponds to
\begin{equation} \label{eq:Sfive}
R_5 \leftrightarrow \frac{g_{5}^2}{8\pi} .
\end{equation}
As one might expect for an S-duality, this is a highly nontrivial relation, involving
all orders in perturbation theory.  At least at first sight, it looks different in $D=5$
than in $D=4$.  

However, it is not essentially different, as we can see by developing
the perturbative expansion for the compactified 5d theory.
To do this, we restrict the momentum along the new $S^1$ to quantized
values
\begin{equation} \label{eq:fivedsum}
\int dp_5\, f(p_5) \rightarrow \frac{1}{R_5} \sum_{n\in\IZ}\, f(\frac{n}{R_5}).
\end{equation}
Taking into account the normalization of the loop integrals, the perturbative
series is actually a series in $g_4^2=R_6/R_5$, just as in $D=4$.  Furthermore,
if we take $R_5\rightarrow 0$ with fixed $R_6/R_5$, we are sending the 5d momentum
states to infinite energy, so with the naive prescription of dropping them, we find
that the relation \eq{Sfive} does reduce to 4d S-duality in the naive way.

This is an example of the regularity of the $R_5\rightarrow 0$ limit which we
referred to in the introduction.  Although it is intuitive that this should work, one
can imagine scenarios in which it did not.  In particular, we might worry about
the consequences of UV divergences in $D=5$.  Such divergences would imply that
the compactified $D=4$ amplitudes get contributions from arbitrarily large
momenta, so that $D=5$ KK states at energies between $1/R_5$ and the cutoff $\Lambda$
would contribute.  In the $(2,0)$ theory, the
S-duality partners of these states are KK states in the compactified $x^6$ direction.
Thus, S-duality requires us to add these states.  This was evident at finite $L$, but if $D=5$
MSYM theory has UV divergences, then we must add them even for arbitrarily small but
non-zero $L$.

To continue the comparison between $D=5$ and $D=4$, let us
rewrite the $D=5$ expansion in terms of $D=4$ parameters,
$$
g_4^2 \equiv \frac{R_6}{R_5}; \qquad L^2 \equiv R_5R_6 .
$$
Now the prescription for compactification turns into
\begin{equation} \label{eq:p-sums}
g_5^2 \int dp_5 \rightarrow g_4^2 \sum_n ; 
\qquad
\frac{1}{p_i^2} \rightarrow \frac{1}{p_{4i}^2 + g_4^2 n_i^2/L^2} .
\end{equation}

The terms with all $n_i=0$ are the $D=4$ amplitudes, while in
principle we could sum the rest of the expansion to identify the finite $L$ corrections
to $D=4$ amplitudes.  Of course, there are subtleties in doing this, analogous
to those in finite temperature computations in conventional gauge theory, but now with the
possibility of additional nonrenormalizable UV divergences.

% Assuming that the $L\rightarrow 0$ limit is regular (it is the $D=4$ theory),

\subsection{Compactified $D=5$ at one loop}

To get a sense of the resulting structures, 
let us consider a one-loop amplitude with $k=4$ external particles. 
In MSYM in any dimension, this amplitude
is the product of a scalar box integral with a kinematic factor,
\begin{equation}
{\CA} = {\CK} \times \sum_{n\in \IZ} I(s,t) .
\end{equation}

Since $p_5^2=n^2 M_5^2$ for a KK mode on $S^1$,
the $D=5$, $k=4$ amplitude is obtained by summing a series of $D=4$ amplitudes
computed with massive box integrals,
\begin{equation}
{\CA}_{D=5} = {\CK} \times \sum_{n\in \IZ} I(s,t;m^2=n^2 M^2_5) .
\end{equation}
An explicit expression for this can be found in 
(\cite{Bern:1995db}, eq. A.8).\footnote{One takes $M=0$ in the result quoted there, which drops the last term.}

As one can verify from the explicit result, because the $m\ne 0$
terms are both UV and IR convergent, one can just scale out $m$, so it
has a non-zero $s=t=0$ limit of order $1/m^4$ and a series expansion
in powers of $s/m^2$ and $t/m^2$ around zero.  Thus the sum will look like
\begin{equation}
\zeta(4)  \frac{1}{M_5^4} +\zeta(6) \frac{s + t}{M_5^6}   + \ldots
\end{equation}
plus the $n=0$ term $\sim 1/st$.

As a function of $M_5$ this is regular at $M_5\rightarrow\infty$, and using
$M_5=g_4/L$ we get a correction
\be
C_8\; L^4 \tr F^4
\ee
to the $D=4$ effective Lagrangian.  This is a sensible ``mock tree order'' term; although it
is generated at one loop, the $g_4$ dependence is cancelled by the $g_4$
dependence of $M_5$.\footnote{Note that this is in conventions
with $F=dA+g_4 A^2$.  To discuss S-duality one would take $S=g_4^{-2}\tr \tilde F^2$ and
$\tilde F=g_4 dA+ g_4^2 A^2$, and  $\tr F^4$ becomes $\tr (\tilde F/g_4)^4$;
this is the convention in \eq{S-dual-amp}.}

One could go on to try to compute an S-dual version of this interaction, by adding in the
$D=6$ KK modes.  A natural guess is that the coefficient is
\be
C_8 L^4 = \sum_{m,n\ne 0} \frac{g_{n_5,n_6}^4}{M_{n_5,n_6}^4} 
\ee
where $g_{n_5,n_6}$ and $M_{n_5,n_6}$ are the effective coupling (to a photon) and mass for a KK mode
with quantized momenta $p_i=n_i/L$.  One can check that
$M^2_{n_5,n_6} = |\tau n_6+n_5|^2$.  While the effective coupling to a 6d KK mode is not known,
under the simple assumption that $g_{n_5,n_6}$ is independent of $n_i$, the sum turns into
a nonholomorphic Eisenstein series,
\be \label{eq:S-dual-amp}
C_8 = \zeta(4) E(\tau,2) \equiv \sum_{m,n\ne 0} \left(\frac{\Im\tau}{|m\tau+n|^2}\right)^2 ,
\ee
and one obtains an S-duality invariant coefficient function.

It would be interesting to predict this term from some string theory argument.  Usually such 
corrections are controlled by $\alpha'$ or $l_{p11}$, but here we dropped this
dependence in defining the $(2,0)$ theory.  On the other hand, one might find
some duality which relates our volume of $T^2$ to these parameters.

The next term in the expansion in $s/m$ and $t/m$ is more
problematic:
\be  \label{eq:one-loop-q}
\frac{L^6}{g_4^2} \tr D^2F^4 .
\ee
While this was a sensible $1/M_5^2$ correction to effective field theory,
it does not make much sense as a term in a perturbative $g_4$ expansion,
and spoils the $g_4\rightarrow 0$ limit at fixed $L$.   What is going on?

In fact, the $g_4\rightarrow 0$ limit at fixed $L$ corresponds to taking 
$R_5\rightarrow\infty$ and $M_5\rightarrow 0$, so it is not surprising that it does not
have a $D=4$ interpretation.  We need to take $L\rightarrow 0$ faster than $g_4\rightarrow 0$
so that $M_5\rightarrow\infty$ to have such an interpretation.  Keeping this in mind, we can
still ask whether particular $D=4$ corrections are regular.

\subsection{Structure of $D=5$ UV divergences}

Let us now consider the $D=4$ effective Lagrangian obtained by integrating out KK modes.
We can apply the same strategy to all orders in the perturbative expansion.  Denote the loop order as $\ell$.
We assume the external momenta
all satisfy $p_i\cdot p_j<<M_5$; then we make the
replacements \eq{p-sums} and integrate out all lines with $p_5\ne 0$,
to get an expression for the corrections to the $D=4$ theory as a sum over $D=4$ loops of
particles with masses $m_a=M_5 n_a$,
\be
\CA_n = \sum_\ell g_4^{2+2\ell} \sum_{n_1,\ldots,n_\ell} \CA_{n,D=4}(p_i;\ m_a) .
\ee
We would then expand this amplitude in external momenta and identify terms as due to operators
in an effective Lagrangian,
\be
\CL = \sum_\Delta C_\Delta \CO_\Delta ,
\ee
where for brevity we label the coefficient of a generic operator with dimension $\Delta$ as $C_\Delta$.

Since the particles in these loops are all massive, the $D=4$ momentum integrals are IR convergent.
Furthermore, at large momentum we can ignore the masses, so since this is MSYM in $D=4$ the
integrals are UV convergent.  Thus we can again estimate them using dimensional analysis, in terms
of the unique scale in the computation, $M_5$.  The only difference with the one loop computation above is that
we can produce more general operators, and the final sum is over $\ell$ independent KK momenta.
Thus,
\be
C_\Delta \sim \sum_\ell g_4^{2+2\ell} \sum_{n_1,\ldots,n_\ell} \frac{1}{(n \cdot M_5)^{\Delta-4}} 
\ee
Of course, the explicit functions of the $n_a$ appearing here will be quite complicated, and
we cannot hope to do these sums explicitly, but we can estimate their dependence on the
cutoff.  If we assume genericity, so that the maximal UV
divergence appears at each order, we find at loop orders $\ell>\Delta-4$
\be
C_\Delta \sim \sum_\ell g_4^{2+2\ell} \left(\frac{\Lambda}{M_5}\right)^{\ell+4-\Delta} \frac{1}{M_5^{\Delta-4}} 
\ee
while at $\ell=\Delta-4$ we find
\be
C_\Delta \sim g_4^{2+2\ell} \log\left(\frac{\Lambda}{M_5}\right) \frac{1}{M_5^{\Delta-4}}
\ee

Now, let us explore the consequences of assuming $\Lambda =1/R_6 = 1/g_4L$.  Recalling that $M_5=g_4/L$, we find
\bea 
C_\Delta &\sim & \sum_\ell g_4^{2\Delta-6} \left(\frac{L}{g_4}\right)^{\Delta-4}  \nonumber\\
 &\sim & L^{\Delta-4} \sum_\ell g_4^{\Delta-2}  \label{eq:power-div}
\eea
from the powerlike divergences, and
\be \label{eq:log-div}
C_\Delta \sim L^{\Delta-4} g_4^{\Delta-2} \log g_4 
\ee
from the log divergence.

Note that the dependence on loop order $\ell$ cancels out of the leading divergences --
the loop counting parameter $g_4^2$ is compensated
by the additional UV divergence $\Lambda/M_5\sim 1/g_4^2$.  
Subleading
divergences at loop order $\ell$ could produce terms with an additional $g_4^2, g_4^6, \ldots$ up to
$g_4^{2\ell-2}$ and $g_4^{2\ell} \log g_4$.

Finally, a finite contribution will go as $g_4^{2\ell+6-\Delta}$.  Although this exponent can be negative
at low loop order, this will be for the same reason as in \eq{one-loop-q}, that the
$g_4\rightarrow 0$ limit at fixed $L$ actually decompactifies, so these are not contributions to the
$D=4$ effective Lagrangian.  Presumably, finite contributions are always regular, by the general
consistency of duality and compactification. 

Turning to actual counterterms, the lowest dimension correction in $D=4$ MSYM is
the $\tr F^4$ term with $\Delta=8$, which only appears at one loop.  At higher loops one can have
counterterms with $\Delta\ge 10$, starting with $\tr D^2 F^4$ which can be generated
logarithmically at $\ell=6$.  This is a superspace D-term \cite{Drummond:2003ex,Berkovits:2009aw} so one expects
that it will be generated, as will operators with $\Delta> 10$ at higher loops,
presumably continuing without bound.

In general, the dependence on the $D=4$ parameters $g_4$ and $L$ 
looks sensible.  $L$ controls the corrections as expected for irrelevant operators, and $g_4$
appears with positive powers, or as log $g_4$ multiplied by a positive power, which at least vanishes
as $g_4\rightarrow 0$.  The problem we observed with the convergent one-loop diagram
does not show up in the UV divergent terms.

\subsection{Discussion and speculations}

The conclusion of this section is that $D=5$ UV divergences
look like they can have a sensible $D=4$ interpretation, under the assumption
that the $(2,0)$ theory provides a cutoff which can be modeled by \eq{cutoff}.
Adding in the subleading divergences and finite terms, we could 
in principle compute corrections to the $D=4$ MSYM low energy limit as
well-defined functions of $g_4$.  To get S-dual results, we must somehow add
in the contibutions of $(2,0)$ KK modes, as in \eq{S-dual-amp}.
One might hope that, instead of
explicitly adding in the contributions of instantonic particles, this could be done by
finding a resummation prescription consistent with modular invariance.

However, since every leading divergence in the loop expansion appears at the same order in $g_4$,
one must ask whether this is a controlled expansion at all.  Clearly this point must be
resolved in order to proceed further.  If the $\ell$ loop term has the generic growth $\ell!$, then it
is hard to see how such an expansion can make sense.  
This might be another argument favoring
UV finiteness of $D=5$ perturbation theory.  Conversely, if there are UV divergences, this might be an argument
that the relation between $(2,0)$ theory and $D=5$ SYM is more complicated than we have assumed,
perhaps involving additional states.

Another possibility is that the coefficients of the cutoff dependent terms do not have the
generic growth $\ell!$.  Perhaps only a finite number of divergences contribute at each order
in the coupling.  We should also remember that the assumption $\Lambda =c/R_6$ is only a model
for some more concrete (and complicated) cutoff provided by
$(2,0)$ theory, such as the loop contributions of
6d KK modes, a form factor, or something else.  For the purpose of computing
corrections to an effective Lagrangian, the cutoff should still
be describable in $D=5$ terms, and perhaps even in terms of a simple ansatz, which
makes the series \eq{power-div} convergent.
For example, the coefficient $c$ could fall off with $\ell$, compensating the $\ell!$.

If the individual corrections \eq{power-div} and \eq{log-div} do make sense, then the
log divergence \eq{log-div}  leads to a very unusual
contribution to the coefficient $C_\Delta$ of a dimension $\Delta$ operator,
proportional to $g_4^{\Delta-2} \log g_4$.  For example, the potential divergence at $\ell=6$ 
is a log divergent coefficient of the operator $L^6 \tr D^2 F^4$.
This would be a very distinctive feature
to try to match from some other construction of $(2,0)$ theory on $T^2$.
Or, perhaps there is some argument that it must vanish, either to satisfy S-duality, or from general
properties of another construction.

Perhaps future developments in loop calculations will tell us whether there are UV divergences,
and provide information enabling us to continue this discussion.  For example,
the planar six loop integrand for the log divergent coefficient of $\tr D^2 F^4$
has been written out explicitly \cite{BCD}. It exhibits some cancellations and is not positive
definite, in contrast to the cases of five or fewer loops in $D = 4 + 6/\ell$.
Numerical integration could help decide whether or not it vanishes. 

\vskip 0.5in
%\vfill
{\it Acknowledgements}

This work originated in a discussion with Lance Dixon at Erice in 2009, and I thank him,
Zvi Bern, and John Joseph Carrasco for discussions and for showing me \cite{BCD}.  I also had
valuable discussions and communications with Nima Arkani-Hamed, Lars Brink, Michael Green, Greg Moore,
Nikita Nekrasov, Nati Seiberg, Kelly Stelle, Andy Strominger and Pierre Vanhove.  
I also thank Neil Lambert for sending me a prepublication copy of \cite{Lambert}.

This research was supported in part by DOE grant DE-FG02-92ER40697.

\end{document}